\documentclass[11pt,fleqn]{article}
\usepackage{epstopdf} %converting to PDF
\usepackage{graphicx}
\usepackage[a4paper,left=2cm,right=2cm]{geometry}
\usepackage{amsmath}
\usepackage{float}
\usepackage[latin1]{inputenc}
\usepackage{amsmath}
\usepackage{amsfonts}
\usepackage{amssymb}
\usepackage{float}
\usepackage[toc,page]{appendix}
\floatstyle{boxed} 
\restylefloat{figure}

\newcommand{\be}{\begin{equation}}
\newcommand{\ee}{\end{equation}}
\newcommand{\bea}{\begin{eqnarray}}
\newcommand{\eea}{\end{eqnarray}}

\begin{document}
\title{Phenomenology of a universal super weak $SU(N)_{w}$ color hypothesis.}
\author{Lobsang Dhargyal \\\\\ (formerly at) Regional Center for Accelerator-based Particle Physics,\\\\\ Harish-Chandra Research Institute, HBNI, Jhusi, Allahabad - 211019, India}
\date{8 Feb, 2020.}

%\preprint{HRI-RECAPP-2018-010}

\maketitle
\begin{abstract}

In this work we will propose a new universal super weak $SU(N)_{w}$ color to which all the SM fermions are assumed to be coupled universally whose confining radius is in the range of the size of galaxies. We will argue that the model has some very interesting consequences such as it could be able to explain the observed rotation velocity anomalies in large galaxies without requirement of specific DM particle. We will also show that it could constitute most of the observed mass of a galaxy, similar like the proton mass which is mostly made up of color interaction dynamics as shown recently, which could explain the observed DM in the rotation velocities of galaxies in galaxy clusters. Also when the N is very small, we will show that it could explain the Big Bang and Hubble's law, Dark Energy, and tentatively also flavor and neutrino oscillation.

\end{abstract}
\maketitle

\section{Introduction.}

The standard-model (SM) of particle physics has been very successful in accounting for the natural phenomenas around us as well as experiments specifically design to test its predictions to very high precisions. However it is known that minimum SM as it stand can not account the small neutrinos masses, dark-matter (DM), dark-energy (DE), why three generations, what are the origin of the flavor of the fermions, strong CP problem, matter-antimatter asymmetry etc. In this work we propose a new super weak $SU(N)_{w}$ color gauge which is coupled to the all fermions of the standard-model (SM) universally and show that it could explain the anomalies observed in the rotation velocities of stars far away from the center of large galaxies as well as the most of the constituent dark-matter (DM) masses of of galaxies which could explain the DM observed in galaxy clusters.

\section{Model details.}
\label{mod-det} 

\begin{figure}[h!]
\begin{minipage}[t]{0.48\textwidth}
\hspace{0.4cm}
\includegraphics[width=2\linewidth, height=8cm]{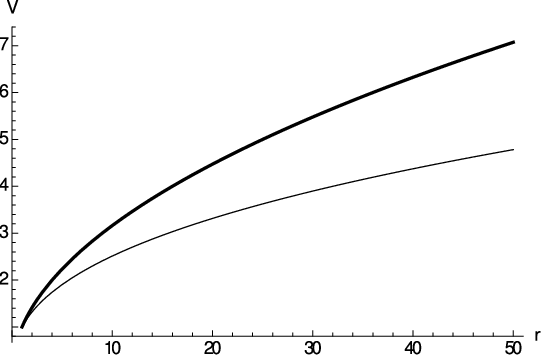}
%%\end{right}
\end{minipage}
\caption{In this plot we have shown $v\ vs\ r$ scaled such that $\frac{K}{mR^{\beta}} = 1$ with $\beta = 0$ (thick dark) and $\beta = -0.2$ (light dark)}
\label{fig1}
\end{figure}

In general Non-Abelian (Yang-Mills (YM)) gauge theories likes of QCD, it has been shown that the coupling strength decreases as distance between the interacting particles decreases (Asymptotic freedom). And also, lattice QCD (LQCD) calculations has strongly indicated that at large distance (strong coupling regime) the QCD potential energy scale as proportional to the distance as $V(r) = Kr$ where K is a constant i.e confinement property is present in QCD. Not only that in QCD numerical stimulations indicated that, unlike in the Abelian $U(1)$ gauge theory case, the transition from the weak coupling regime (short distance) to the strong coupling regime (long distance) is smooth without any indication of a phase transition like behavior, see \cite{ChengLi} and \cite{PS}. In the QCD like YM theories the coupling scaling with energy is expressed as \cite{ChengLi}\cite{PS}\cite{HM}
\be
\alpha_{s}(Q^{2}) = \frac{\alpha_{s}(\mu)}{1 + \frac{1}{4\pi}(\frac{11N}{3}-\frac{2}{3}n_{f})\alpha_{s}(\mu)\log{\frac{Q^{2}}{\mu^{2}}}}
\ee
with $\alpha_{s}(\mu^{2}) = \frac{g^{2}_{s}}{4\pi}$ being the coupling strength at scale $\mu^{2}$ and as is clear from the above equation for $Q^{2} > \mu^{2}$ as $Q^{2}$ increases, the coupling $\alpha_{s}(Q^{2})$ decreases to indicate the asymptotic freedom, where $n_{f}$ is the number of quark flavors included in the loop calculations.\footnote{As such it turns out that the SM fermions can only have SU(2) new weak color symmetry. The following arguments are true provided we take a dark sector fermions with at least one of the dark fermions carrying a small (mili-charge) EM charge which itself carrying a dark strong color similar like the SM strong force which can force it to be physical only in the form of dark mesons, dark baryons etc. In this case the fermions them self can be very light and still their color neutral manifestations could be pretty heavy similar to the SM mesons and baryons which can explain why they are not produced in abundance at colliders even though the fermions are light if some of the dark fermions carry EM mili-charges. However a mili-charged light dark fermion can induce scattering between EM light and this new dark weak color field back ground with very small energy transfer, via box loop and higher loops, which can give rise to a red shift similar to Hubble's law. One such observation could be the reported X17 boson by ATOMKI experiment, which in this case could be a neutral dark meson formed by the mili-charged dark fermions which are produced on shell via a virtual photon which decays into an electron-positron pair via a virtual photon. (In general where box diagram is non-zero; e.g in high energy SM EM interaction as well as model presented in this paper, a photon can decay into three massless particles if the three final particles are all emitted co-linear to the direction of the initial photon and also co-linear decay of photon into a photon and massless particles (e.g gravitons) for instance can give rise to Hubble like law. This work is probably first to point out that Photons co-linearly decaying into massless final particles is allowed; especially into gravitons is to be noted.)} Though this formula is based on the perturbation analysis, which will breakdown when the coupling is no longer small, still $\mu$ turn out to be a very useful energy scale at which the coupling become large.  Taking inspirations from these two key properties of Non-Abelian gauge theories of asymptotic freedom at short distances and confinement at large distances, we propose a new Universal Super Weak $SU(N)_{w}$ color local gauge which couples to all the fermions of SM universally and whose confining radius is in the order of largest galaxies.\footnote{in general it could be any gauge group, not necessarily SU(N), that has the above mensioned features.... even not from any gauge forces with feature that it reaches a constant magnitude at certain large distance scale will do....}
Now given that we assume the coupling strength is very weak, it's confining regions will be much larger than the size of the proton. Here we take the coupling strength such that the confining region of the new $SU(N)_{w}$ is in the range of the size of galaxies. In this case due to asymptotic freedom in the super weak coupling regions or super short distances compared to the confining scale (scale of largest galaxies), i.e at the scale of solar system and smaller, the new $SU(N)_{w}$ force is much weaker than gravity so that we can ignore its presence.\footnote{...to actually calculate we need to know the flux of the weak gluons, but we would like to point out here that, qualitatively we can understand why object do not slowed by new gluons is that although gluons with higher momentum is coming from the front than from behind but since ones that are from front are of higher energy than ones from behind, front gluon flux are less coupled than the gluon flux from behind..........side note: can a CMB like back ground due to new U(1) gauge contribute to an object's inertia?} But as the distance between two objects charged under $SU(N)_{w}$ grows, the coupling strength grows and eventually it surpasses strength of gravity and at the scale of galaxies lets say it goes into non-perturbation regime. In general it has been verified from lattice calculations that in the non-perturbation regime, the potential energy between two objects charged under the new $SU(N)_{w}$ grows as $V(r) = kr$ and so the force between them will grow as $f\hat{r} = k\hat{r}$ where $\hat{r}$ is unit vector in the direction of r and k a constant. So the total force acting on an object very far away from the center of a galaxy will be given by $F\hat{r} = \sum_{i} f\hat{r_{i}} = K(\frac{r}{R})^{\beta}\hat{r}$ where $(\frac{r}{R})^{\beta}$ is the r (here on r denote the radial distance from the center of the galaxy to the test star) dependence induced by integral over the density of matter which are charged under the new $SU(N)_{w}$ and R is the confining radius with $\beta$ being a parameter to be fitted with the galaxy density distribution.\footnote{note that in reality the galaxy mass distribution profile could be more complicated than the simple one stipulated here......} Now it is known that the key problem with reported anomaly in rotation curves of stars in large galaxies is that it reaches almost a constant value while the Newtonian gravity (which is very good approximation of the general relativity in weak field regimes) predicts a sharp decline in the velocity as the test star is further and further away from the center of the galaxy. Here we would like to point out that this new $SU(N)_{w}$ will be able to explain this phenomena. Assuming that we are in a region where contribution to the rotational velocity due to gravity is negligible, then we have $m\frac{v^{2}}{r}\hat{r} = K(\frac{r}{R})^{\beta} \hat{r}$ so $v = \sqrt{\frac{K}{mR^{\beta}}}\sqrt{r^{1+\beta}}$. In Figure (\ref{fig1}) we have shown $v\ vs\ r$ scaled such that $\frac{K}{mR^{\beta}} = 1$ with $\beta = 0$ (thick dark) and $\beta = -0.2$ (light dark).
From the Figure (\ref{fig1}) it is clear that for large $\beta$ with $\beta\ < 0$, the graph becomes flatter and flatter which seems to be required from the observed rotation curves of stars far away from the center in large galaxies. In general for a star with given mass (m), there are three free parameters $K$, $R$ and $\beta$ to fit with the data from the rotation velocity and density distribution measurements.\\
\\
Now it is known that the first hint of DM is from the rotation velocity anomalies in large galaxies but as shown above the main course of this anomaly may be attributable to a new $SU(N)_{w}$. Also as shown in a recent calculation in \cite{YBYang}, that most of the mass of the proton comes from the quark energy and glue field energy contributing about 32\% and 36\% respectively with about 23\% coming from the trace anomaly and the remaining about 9\% contribution from the quark scalar condensates with less than 1\% due to the valence quark masses (due to Higgs boson). Since proton is a confine state of quarks and gluons, similarly in our new $SU(N)_{w}$ the whole galaxy being a confine state of matter charged under the new $SU(N)_{w}$, we also expect that most of the galaxy masses comes from the condensates of the constituents. Then similar to the string analogy of QCD confinement, we can say that most of the mass of a galaxy could be stored in the string tensions of the confining new super weak color mediator gluons between the constituents of the galaxy which are charged under the new $SU(N)_{w}$. In this sense we may be able to attribute the DM effects observed in the galaxy clusters to energy stored in this string tensions of the new super weak color mediator gluons inside each galaxy in the galaxy cluster\footnote{Also we would like to point out that an easy explanation of the existence of large amount of X-ray emitting gas in the inter-galactic region in galaxy clusters is to assume that there exist some light DM particles and only these new particles are charged under the new $SU(N)_{W}$ with confinement region at the scale of galaxies, then it is expected that these particles will not form large structures if they are very light and so will be able to explain the DM halos without observations of dark stars, dark planets etc (this model of DM will work even for the bosons being SU(N) local gauge bosons)...... on the other hand the galaxy clusters could be like the multi-nucleonic atomic nucleus with inter-galactic gases could be like not yet condensed galaxies or other exotic states such as quark-gluon plasma, or even some kind of meson like states with galaxy clusters bound together by Yukawa like force etc....... in general a weak gluon and anti-gluon can form a color singlet scalar or a vector or a spin 2 bosons which could be long ranged due to their expected small masses...}. These are only tentative arguments and analogies, as to actually show how much of the mass of a galaxy could be attributed to this string tension we need to know the confining radius and the coupling constant at a particular scale etc., which has not been fitted yet. We would like to point out here that if this is the force that actually binds the galaxies together, then we can actually study much about galaxies collisions from proton proton collisions at LHC and vice versa! Also it seems our model can explain the mechanism behind the spiral arms of spiral galaxies from the angular pressure distributions that have been recently worked out, see the diagram on page 20 of \cite{MVP}.\\
\\
Also we would like to point out a very interesting special case of this hyper weak gauge coupling regime here, that is the case when N is very small such as 2, 3, 4, etc., then it turn out the confining region is much larger than the horizon of about 14000 mpc (the observable patch) for the new force to be much weaker than gravity at 1 GeV scale. However an interesting consequences of $SU(N)_{W}$ with low N for large scale cosmos is that suppose a perturbation to such an object has happened in some early time, may be due to collision with another object like it, such that the observable patch around our galaxy has been pushed towards the center, then first an equilibrium in the  weak gluon field get restored which sets up a potential gradient in the form of a radially outward-directed force mountain \cite{MVP} along with sound waves with very large wave length in the weak gluon field. Now if the matter in our patch of the observable cosmos is climbing (up or down) this force mountain along with coming out of a very concentrated region in the past into a rarefying region, due to sound wave propagation through our observable patch, then the Big Bang can be taken as a moment in the past when this sound wave concentrated large chunk of the cosmos near our observable patch into a very small region which eventually get rarefied (which we call Big Bang) with the observable patch climbing the force mountain seen from our galaxy will appear in the form of the Hubble's law with smooth but increasing slope of the pressure mountain appearing as accelerated expansion which we interpreted as dark energy (DE) or DE could be due to combination of sound wave and gravity effect.\footnote{Also present way of looking at cosmological data indicates that we are in a region where angular velocity is much smaller than the radial velocity....}\\
There is another possible source of the red shift of light from distance objects, and that is due to the inelastic scattering of the photons with the much colder CMB background or the weak gluon field background (via the box loop diagram)\footnote{for instance with scattering of photon and weak guon field via box loop with final offshell weak gluon decaying into two weak gluons one of which splits into 2 more guons etc. or even neutrino anti-neutrino pairs, i.e $\gamma + g_{w} \rightarrow \gamma + g_{w}^{*} \rightarrow g_{w} + (g_{w}^{*} \rightarrow g_{w}/\nu + g_{w}/\bar{\nu})$} and transfer a very tiny amount of energy and momentum to the tail end of weak gluon field or CMB distribution at each collision, then photons emitted from farther away sources are expected to be more red shifted then nearer ones and so roughly the red shift should be proportional to the distance just as Hubble's law requires. The observed red shift could probably be due to combinations of this and other factors pointed out in the preceding paragraph, however if the observed red shift is entirely due to this inelastic scattering then the red shift does not implies expanding universe. We would also like to point out that, what ever the source of the CMB, if the CMB photons are emitted very long time ago then it is possible that it may have reached equilibrium with the much colder back ground weak gluon field (WGF), in that case the CMB temperature is also the temperature of the WGF or it could be that CMB is in the process of reaching equilibrium with the WGF but not yet reached, then the temperature of the WGF should be much colder than the presently measured temperature of CMB.\footnote{another interesting observation we made about small N is that, when N = 3, it seems it can tentatively explain why there are three flavor by attributing flavors as frozen colors as well as neutrino oscillations as scattering process (and decays) of neutrinos with these back ground WGF and change it's color/flavor..... in that case, the coupling strength of this new force relative to EW coupling can be estimated from neutrino oscillation......}

\section{Conclusions.}
\label{sect:conclusions}
In this work we have proposed a new universal super weak $SU(N)_{w}$ color to which all the SM fermions are assumed to be coupled universally whose confining radius is in the range of the size of galaxies. We have argued that the model has some very interesting consequences such as it could be able to explain the observed rotation velocity anomalies in large galaxies without requirement of specific DM particle. We have also shown that it could constitute most of the observed mass of a galaxy, similar like the proton mass which is mostly made up of color interaction dynamics as shown recently, which could explain the observed DM in the rotation velocities of galaxies in galaxy clusters. Also we have shown that the weak gauge coupling $SU(N)$ with small N could potentially explain Big Bang, Hubble's law, Dark energy, and tentatively also flavor and neutrino oscillation.

%\newpage

{\large Acknowledgments: \large} This work was partially supported by funding available from the Department of Atomic Energy, Government of India, for the Regional Center for Accelerator-based Particle Physics (RECAPP), Harish-Chandra Research Institute.


\begin{thebibliography}{99}

\bibitem{YBYang} Y. B. Yang, J. Liang, Y. J. Bi, Y. Chen, T. Drape, K. F. Lui and Z. Lui, \textsl{PRL 121, 212001 (2018)}

\bibitem{ChengLi} T. P Cheng and L. F. Li, \textsl{Gauge theory of elementary particle physics. Oxford university press.}

\bibitem{PS} M. E. Peskin and D. V. Schroeder, \textsl{An introduction to quatum field theory. Westview press.}

\bibitem{MDS} Mathew D. Schwartz, \textsl{Quantum Field Theory and the Standard Model. Cambridge University Press.}

\bibitem{HM} Francis Halzen and Alan D. Martin, \textsl{Quarks and Leptons : An Introductory Course in Modern Particle Physics. John Wiley and Sons, Inc.}

\bibitem{MVP} M. V. Polyakov and P. Schweitzer, \textsl{arXiv: 1805.06596v3 [hep=hp] 14 Sep 2018.}


\end{thebibliography}
\end{document}